\documentclass[aps,showpacs,preprintnumbers,amsmath,amssymb]{revtex4}
\usepackage{dcolumn}
\usepackage[dvips]{epsfig}

\usepackage{float}
\usepackage{graphics,epsfig}
\usepackage{graphicx}
\usepackage{dcolumn}
\usepackage{bm}
\usepackage{gensymb}
\usepackage{subfigure}
\usepackage{diagbox}
\usepackage{enumerate}

\begin{document}
\baselineskip=0.5 cm
\title{\bf Determination of the spin parameter and the inclination angle by the relativistic images in black hole image}

\author{ Mingzhi Wang$^{1}$\footnote{Corresponding author: wmz9085@126.com}, Songbai Chen$^{2,3}$\footnote{csb3752@hunnu.edu.cn}, Jiliang Jing$^{2,3}$\footnote{jljing@hunnu.edu.cn}
}
\affiliation{$ ^1$School of Mathematics and Physics, Qingdao University of Science and Technology, Qingdao, Shandong 266061, People's Republic of China \\
$ ^2$Institute of Physics and Department of Physics, Key Laboratory of Low Dimensional Quantum Structures
and Quantum Control of Ministry of Education, Synergetic Innovation Center for Quantum Effects and Applications,
Hunan Normal University, Changsha, Hunan 410081, People's Republic of China\\
$ ^3$Center for Gravitation and Cosmology, College of Physical Science and Technology,
Yangzhou University, Yangzhou 225009, China}

\begin{abstract}
\baselineskip=0.4 cm

{\bf Abstract} We studied the relativistic images caused by strong gravitational lensing in Kerr black hole images, which carry some essential signatures about the black hole space-time. We defined a new celestial coordinates whose origin is the center of black hole shadow to locate the relativistic images. Under the influences of the dragging effect caused by rotating black hole and the inclination angle of observer, the relative positions between the primary and secondary images are different with the different Kerr spin parameter $a$ and the observer's inclination angle $i$, so it can be used to determine the value of $a$ and $i$. We provided the specific approach to measure the value of $a$ and $i$ by the relativistic images. The time delays between the primary and secondary images are different with the different $a$ and $i$. The time delays in conjunction with the relative positions between the primary and secondary images could allow us to measure the value of $a$ and $i$ more precisely. These relativistic images are as unique as fingerprints for black hole space-time, by which one can further determine other parameters of all kinds of compact objects and verify various theories of gravity. Our results provide a new method to implement parameter estimation in the study of black hole physics and astrophysics.

\end{abstract}

 \maketitle
\newpage
\section{Introduction}

Nowadays, Event Horizon Telescope (EHT) Collaboration et al have revealed the first images of the supermassive black holes at the centre of the giant elliptical galaxy M87 \cite{eht,fbhs1,fbhs2,fbhs3,fbhs4,fbhs5,fbhs6} and the Milky Way Galaxy\cite{sga1,sga2,sga3,sga4,sga5,sga6}. It opens a new era in the fields of astrophysics and black hole physics, and attracts an increasing number of researchers into the study of black hole images. A very important element in black hole image is black hole shadow, a dark silhouette\cite{synge,sha2,lumi,sha3}. The dark shadow appears because the light rays close to black hole are captured by black hole, thereby leaving a black shadow in the observer's sky. Black hole shadow can carry many information of the space-time geometry around the compact object, so it plays a vital role in the study of black holes (constraining black hole parameters)\cite{sha9,sha8,dressed,Intcur,bhparam,obsdep,constr}, probing some fundamental physics issues including dark matter\cite{polar7, drk, polar8,shadefl,shasgra} and verification of various gravity theories\cite{safeg,lf,sha10,fR, 2101, 2107, 2111}. Other researches on black hole shadows have been investigated in Refs.\cite{fpos2,sb10,sw,swo,astro,chaotic,zhengwen23,sha18,binary,bsk,my,sMN,swo7,mbw,mgw,schm,scc,sha4,sha5,sha6,sha7,sha11,sha111,sha12,sha13,sha14,who,whr,sha141,sha15,sha16,sb1,sha17,sha19,sha191,sha192,sha193,sha194,shan1,shan1add,shan3add,rr,pe,halo,review,lf2,Zeng2020vsj,Zeng2020dco,lens,knn,bieu,ssp,nake,nakeop,BI}. It is hope that these information imprinted in black hole shadow can be captured in the future astronomical observations including the upgraded Event Horizon Telescope and BlackHoleCam\cite{bhc}.

It is widely believed that a black hole at the center of the galaxy would possess a spin. A neutral rotating black hole is well described by the Kerr metric with two parameters, the mass $M$ and the spin parameter $a$. Kerr black hole shadow gradually becomes a D-shaped silhouette with the increase of spin parameter\cite{sha2}. Furthermore, it also depends on the inclination angle between the observer's line of sight and the spin axis of black hole. Hioki and Maeda\cite{sha9} calculated the observable radius $R_{s}$ and the deviation parameter $\delta_{s}$ of Kerr black hole shadow to determine the spin parameter $a$ and the observer's inclination angle $i$. These observables also can be used on measuring other parameters of Kerr-like black hole\cite{sha8,dressed,Intcur,bhparam,obsdep} and testing the other alternative theories of gravity\cite{safeg}.

When observing a black hole, there may be relativistic images around the black hole shadow. The relativistic images are the images of luminous celestial objects caused by the strong gravitational lensing\cite{sglensing}. There are two infinite sets of relativistic images of one light source on either side of the black hole shadow. But only two of them are easily detected, namely the primary image (PI) and secondary image (SI), and the other relativistic images are concentrated near shadow boundary, can not be distinguished. Not only black hole shadow but also the relativistic images can carry some important signatures about the black hole space-time. The relative position between PI and SI in Kerr black hole images will be influenced by the dragging effect and the inclination angle. These relativistic images are as unique as fingerprints for black hole space-time, by which one can further determine parameters of all kinds of compact objects, such as Kerr-like black holes and so on, and verify various theories of gravity. Our results could provide a new method to implement parameter estimation in the study of black hole physics and astrophysics.

The paper is organized as follows. In section II, we briefly introduce the celestial coordinates to locate the coordinates of the image points in observer's sky, and the observables of Kerr black hole shadow to determine $a$ and $i$. In section III, we calculated four couples of PI and SI around black hole shadow with different $a$ and $i$, and determined $a$ and $i$ by the relative positions between PI and SI. In section IV, we calculated the time delays between PI and SI, which can help to determine the value of $a$ and $i$ more precisely. Finally, we present a conclusion. In this paper, we employ the geometric units $G=c=M=1$.

\section{Determination of the Kerr spin parameter and the inclination angle by Kerr black hole shadow}
The Kerr metric describes a neutral rotating black hole specified by two parameters, the mass $M$ and the spin parameter $a$. In the Boyer-Lindquist coordinates, the Kerr metric has a form
\begin{eqnarray}
\label{kerr}
ds^{2}=-(1-\frac{2Mr}{\rho^{2}})dt^{2}+\frac{\rho^{2}}{\Delta}dr^{2}+\rho^{2} d\theta^{2} +\sin^{2}\theta\bigg(r^{2}+a^{2}+\frac{2Mra^{2}\sin^{2}\theta}{\rho^{2}}\bigg)d\phi^{2}
-\frac{4Mra\sin^{2}\theta}{\rho^{2}}dtd\phi,
\end{eqnarray}
where
\begin{equation}
\Delta=a^{2}+r^{2}-2Mr,\;\;\;\;\;\;\;\;\;\; \rho^{2}=r^{2}+a^{2}\cos^{2}\theta.
\end{equation}
The event horizon of Kerr black hole is at $r_{h}=M\pm\sqrt{M^{2}-a^{2}}$, only exists for $|a|\leq M$. The lights entering the event horizon are absorbed by black hole, thus a black hole shadow emerges. But it is the photon sphere that determines the boundary of black hole shadow. The photon sphere is composed of unstable photon spherical orbits which satisfy
\begin{eqnarray}
\label{rr1}
\dot{r}=0, \;\;\;\;\;and \;\;\;\;\;\ddot{r}=0.
\end{eqnarray}
The light rays that enter the photon sphere will be captured by the black hole; the light rays that fly by the photon sphere could reach infinity; the light rays that spiral asymptotically towards the photon sphere constitute the black hole shadow boundary.

In observer's sky, one can use the celestial coordinates to define the coordinates of the image points in black hole images. In Refs.\cite{my,sMN,lf,mbw,mgw,scc,pe,halo,review}, we calculated the celestial coordinates in axially symmetric space-times as
\begin{eqnarray}
\label{ccxd}
&&x=-r\frac{p^{\hat{\phi}}}{p^{\hat{r}}}|_{(r_{o},\theta_{o})}, \nonumber\\
&&y=r\frac{p^{\hat{\theta}}}{p^{\hat{r}}}|_{(r_{o},\theta_{o})},
\end{eqnarray}
where $p^{\hat{\mu}}$ denotes the photon's four-momentum, which was measured locally by the observer at ($r_{o}, \theta_{o}$). The locally measured four-momentum $p^{\hat{\mu}}$ can be expanded by the four-momentum $p^{\mu}$ of photon as a form \cite{sha2,sw,swo,astro,chaotic,my,sMN,lf,swo7,mbw,mgw,scc}
\begin{eqnarray}
\label{kmbh}
p^{\hat{t}}&=&\sqrt{\frac{g_{\phi \phi}}{g_{t\phi}^{2}-g_{tt}g_{\phi \phi}}} E-\frac{g_{t\phi}}{g_{\phi\phi}}\sqrt{\frac{g_{\phi \phi}}{g_{t\phi}^{2}-g_{tt}g_{\phi \phi}}}L_{z}, \nonumber\\
p^{\hat{r}}&=&\frac{1}{\sqrt{g_{rr}}}p_{r},\;\;\;\;\;
p^{\hat{\theta}}=\frac{1}{\sqrt{g_{\theta\theta}}}p_{\theta},\;\;\;\;\;
p^{\hat{\phi}}=\frac{1}{\sqrt{g_{\phi\phi}}}L_{z},
\end{eqnarray}
where $E$ and $L_{z}$ are the two conserved quantities in photon motion, i.e. the energy and the z-component of the angular momentum, expressed as
\begin{eqnarray}
\label{EL}
E=-p_{t}=-g_{tt}\dot{t}-g_{t\phi}\dot{\phi},\;\;\;\;\;\;\;\;\;
L_{z}=p_{\phi}=g_{\phi\phi}\dot{\phi}+g_{\phi t}\dot{t}.
\end{eqnarray}

The image points in the boundary of black hole shadow correspond to the light rays that spiral asymptotically towards photon sphere. And due to the observer is far away from black hole, $r_{o}$ take the limit $r_{o}\rightarrow\infty$. The analytic expressions for the boundary of Kerr black hole shadow in the celestial coordinates are\cite{sha2,sw,swo,astro,chaotic,my,sMN,lf,swo7,mbw,mgw,scc}
\begin{eqnarray}
\label{xd1w}
x&=&-\frac{\eta}{\sin \theta_{o}}, \nonumber\\
y&=&\pm\sqrt{\sigma+a^{2}\cos^{2}\theta_{o}-\eta^{2}\cot^{2}\theta_{o}},
\end{eqnarray}
where the impact parameter $\eta$ and $\sigma$ are conserved quantities in the photon motion,
\begin{eqnarray}
\eta&=&\frac{L_{z}}{E}=-\frac{r^{2}(r-3M)+a^{2}(r+M)}{a(r-M)}, \nonumber \\
\sigma&=&\frac{r^{3}[4a^{2}M-r(r-3M)^{2}]}{a^{2}(r-M)^{2}}.
\label{cs}
\end{eqnarray}

\begin{figure}[ht]
\center{\includegraphics[width=8cm ]{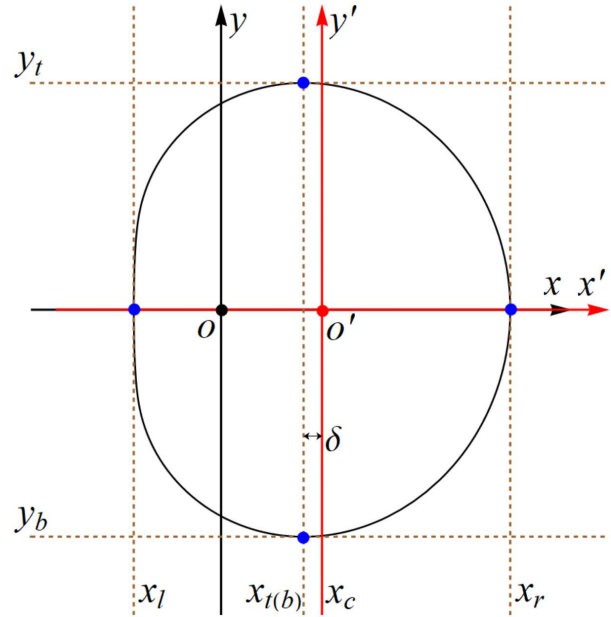}
\caption{The leftmost point ($x_{l}$, $y_{l}$), the rightmost point ($x_{r}$, $y_{r}$), the topmost point ($x_{t}$, $y_{t}$), the bottommost point ($x_{b}$, $y_{b}$) and the center ($x_{c}, y_{c}$) $=$ ($\frac{x_{l}+x_{r}}{2},\frac{y_{t}+y_{b}}{2}$) of Kerr black hole shadow($a=0.998M, i=90^{\circ}$). The distortion $\delta$ is defined as $x_{c}-x_{t(b)}$. The new celestial coordinates ($x', y'$) take the center ($x_{c}, y_{c}$) as the origin.}
\label{xzb}}
\end{figure}

\begin{figure}[ht]
\center{\includegraphics[width=15cm ]{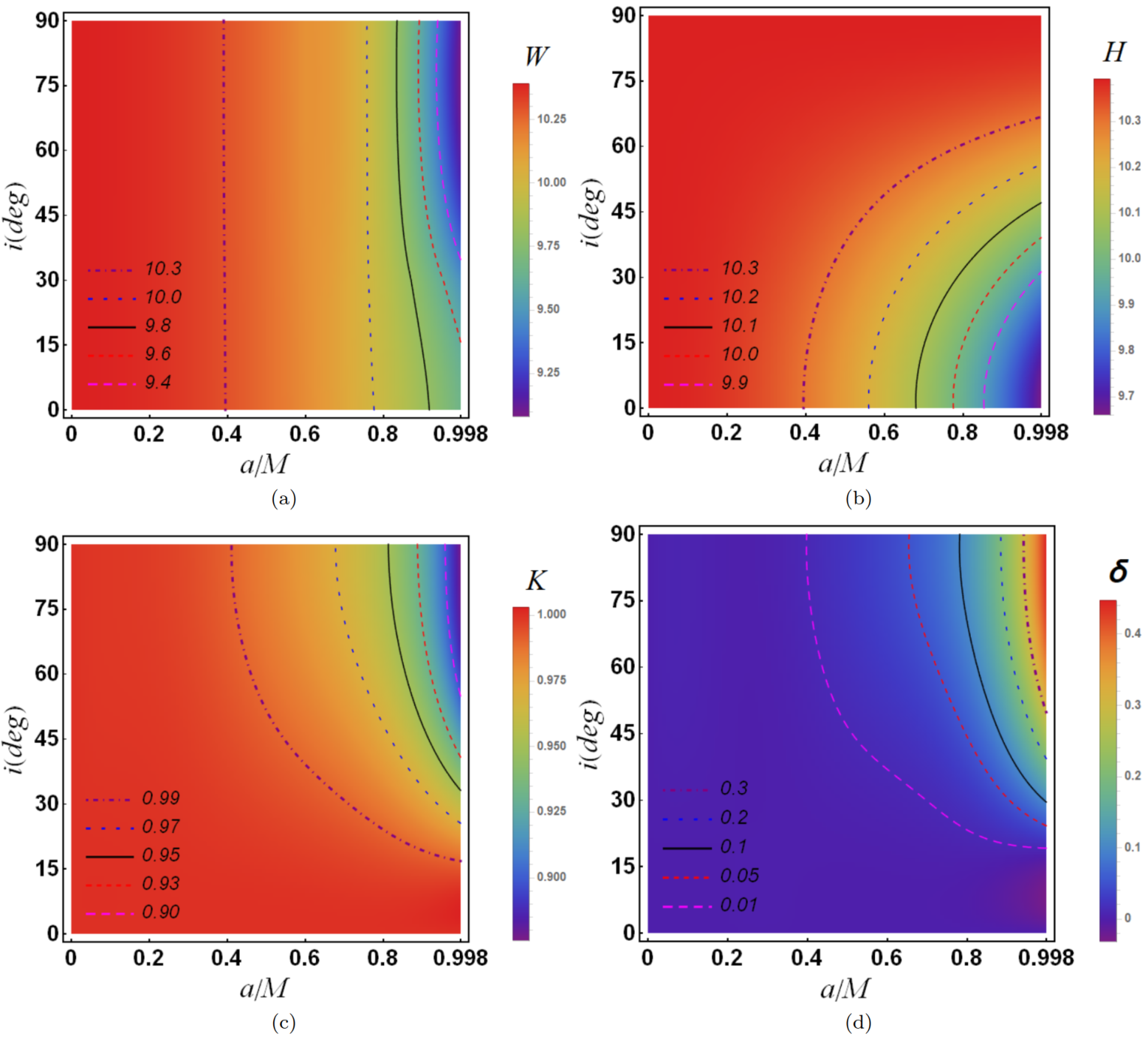}
\caption{The contour maps of the width $W$, height $H$, oblateness $K$ and distortion parameter $\delta$ of Kerr black hole shadows in terms of the spin parameter $a$ and inclination angle $i$.}
\label{whkd}}
\end{figure}

It is shown that Kerr black hole shadow depends on both the spin parameter $a$ and the observer's inclination angle $i$ (i.e. $\theta_{o}$) in Ref.\cite{sha2}. Thus $a$ and $i$ can be determined by calculating the shadow's observable radius $R_{s}$ and the deviation parameter $\delta_{s}$ describing the deviation of the shadow from a circle\cite{sha9}. We calculated several additional observables for Kerr black hole shadow that might be used to quantify $a$ and $i$ more easily. In Kerr black hole shadow with $a=0.998M, i=90^{\circ}$, Fig.\ref{xzb}, one can easily find four characteristic points: the leftmost point ($x_{l}$, $y_{l}$), the rightmost point ($x_{r}$, $y_{r}$), the topmost point ($x_{t}$, $y_{t}$) and the bottommost point ($x_{b}$, $y_{b}$) of the shadow. TABLE \ref{tabl}, \ref{tabr} and \ref{tabsx} exhibit the coordinates of ($x_{l}$, $y_{l}$), ($x_{r}$, $y_{r}$) and ($x_{t}$, $y_{t}$) respectively for Kerr black hole shadow under the different $a$ and $i$ in celestial coordinates. For the bottommost point ($x_{b},y_{b}$)$=$($x_{t},-y_{t}$). In addition, we define the shadow center ($x_{c}, y_{c}$) $=$ ($\frac{x_{l}+x_{r}}{2},\frac{y_{t}+y_{b}}{2}$), and the coordinates ($x_{c}, y_{c}$) with different $a$ and $i$ are shown in TABLE \ref{tabc}. One can find both the $x_{c}$ and $x_{t(b)}$ increase as $a$ and $i$ increase with the influence of the dragging effect. There is a distortion in Kerr black hole shadow, that is $x_{c} \neq x_{t(b)}$. Using these point coordinates, we define four black hole shadow's observables: the width $W=x_{r}-x_{l}$, the height $H=y_{t}-y_{b}$, the oblateness $K=W/H$ and the distortion $\delta=x_{c}-x_{t(b)}$. Fig.\ref{whkd} shows the contour maps of the width $W$, height $H$, oblateness $K$ and distortion $\delta$ of Kerr black hole shadow in terms of the spin parameter $a$ and inclination angle $i$. As $a$ increases, the width $W$ decreases due to the dragging effect, as shown in Fig.\ref{whkd}(a). The width $W$ is almost independent to the inclination angle $i$ when $a<0.9M$, and decreases as $i$ increases when $a>0.9M$. This is because the D-shaped silhouette shadow emerges when $a>0.9M$ and $i$ gradually converges to $90^{\circ}$. From Fig.\ref{whkd}(b), one can find the height $H$ of shadow decreases as $a$ increases, and increases as $i$ increases. Especially, the width $H$ is independent to $a$ when $i=90^{\circ}$, but decreases as $a$ increases when $i=0^{\circ}$. From Fig.\ref{whkd}(c), one can find no matter what values $a$ and $i$ take (except $a=i=0$), the oblateness $K$ is always less than $1$, and it decreases as $a$ and $i$ increase. From Fig.\ref{whkd}(d), one can find the distortion $\delta$ of shadow increases as $a$ and $i$ increase, which means both $a$ and $i$ can manifest the distortion of Kerr black hole shadow. For a given $a$ and $i$, only a set of values of $W$, $H$, $K$, and $\delta$ can be found in Fig.\ref{whkd}. After locating the four characteristic points of black hole shadow, one can measure the width $W$, height $H$, oblateness $K$ and distortion $\delta$, and then determine $a$ and $i$ by use of the one-to-one correspondence between ($W$, $H$, $K$ and $\delta$) and ($a$ and $i$), as shown in Fig.\ref{whkd}.

For the case of the spin parameter $a<0.4M$, the changes of Kerr black hole shadows with $a$ and $i$ are very little, so neither the observables $W$, $H$, $K$, $\delta$ of shadow we proposed nor the observables $R_{s}$, $\delta_{s}$ proposed in Ref.\cite{sha9} can determine $a$ and $i$ very accurately. What's more, measuring the distortion of black hole shadows requires very high accuracy of astronomical telescope that the current EHT does not have yet. So better measurements or more visible observables are needed to determine the Kerr spin parameter $a$ and the inclination angle $i$.

\begin{table}[ht]
\begin{center}
\begin{tabular}{|c|c|c|c|c|c|c|c|c|c|c|}
\hline \hline  \diagbox {$a/M$}{$i(deg)$} & $0^{\circ}$&$30^{\circ}$ &$45^{\circ}$&$60^{\circ}$&$90^{\circ}$ \\
\hline
 $0.2$ & (-5.185,0)& (-4.984,0)& (-4.901,0)&(-4.837,0)& (-4.783,0)\\
 \hline
 $0.4$ & (-5.149,0)& (-4.741,0)& (-4.573,0)&(-4.445,0)& (-4.337,0)\\
 \hline
 $0.6$ & (-5.085,0)& (-4.455,0)& (-4.197,0)&(-4.001,0)& (-3.838,0)\\
 \hline
 $0.8$ & (-4.985,0)& (-4.094,0)& (-3.729,0)&(-3.457,0)& (-3.237,0)\\
 \hline
 $0.998$ & (-4.830,0)& (-3.511,0)& (-2.868,0)&(-2.412,0)& (-2.111,0)\\
\hline\hline
\end{tabular}
\end{center}
\caption{The leftmost coordinates ($x_{l}, y_{l}$) of Kerr black hole shadows with the spin parameter $a/M=0.2, 0.4, 0.6, 0.8, 0.998$ and the inclination angle $i=0^{\circ}, 30^{\circ}, 45^{\circ}, 60^{\circ}, 90^{\circ}$ in the celestial coordinate system ($x,y$).}\label{tabl}
\end{table}
\begin{table}[ht]
\begin{center}
\begin{tabular}{|c|c|c|c|c|c|c|c|c|c|c|}
\hline \hline  \diagbox {$a/M$}{$i(deg)$} & $0^{\circ}$&$30^{\circ}$ &$45^{\circ}$&$60^{\circ}$&$90^{\circ}$ \\
\hline
 $0.2$ & (5.185,0)& (5.385,0)& (5.468,0)&(5.532,0)& (5.586,0)\\
 \hline
 $0.4$ & (5.149,0)& (5.555,0)& (5.722,0)&(5.850,0)& (5.958,0)\\
 \hline
 $0.6$ & (5.085,0)& (5.708,0)& (5.961,0)&(6.154,0)& (6.316,0)\\
 \hline
 $0.8$ & (4.985,0)& (5.844,0)& (6.188,0)&(6.447,0)& (6.662,0)\\
 \hline
 $0.998$ & (4.830,0)& (5.963,0)& (6.401,0)&(6.727,0)& (6.997,0)\\
\hline\hline
\end{tabular}
\end{center}
\caption{The rightmost coordinates ($x_{r}, y_{r}$) of Kerr black hole shadows with the spin parameter $a/M=0.2, 0.4, 0.6, 0.8, 0.998$ and the inclination angle $i=0^{\circ}, 30^{\circ}, 45^{\circ}, 60^{\circ}, 90^{\circ}$ in the celestial coordinate system ($x,y$).}\label{tabr}
\end{table}

\begin{table}[ht]
\begin{center}
\begin{tabular}{|c|c|c|c|c|c|c|c|c|c|c|}
\hline \hline  \diagbox {$a/M$}{$i(deg)$} & $0^{\circ}$&$30^{\circ}$ &$45^{\circ}$&$60^{\circ}$&$90^{\circ}$ \\
\hline
 $0.2$ & (0, 5.185)& (0.201, 5.187)& (0.283, 5.190)&(0.347, 5.193)& (0.400, 5.196)\\
 \hline
 $0.4$ & (0, 5.149)& (0.406, 5.161)& (0.571, 5.173)&(0.696, 5.185)& (0.800, 5.196)\\
 \hline
 $0.6$ & (0, 5.085)& (0.620, 5.114)& (0.867, 5.142)&(1.050, 5.170)& (1.200, 5.196)\\
 \hline
 $0.8$ & (0, 4.985)& (0.854, 5.044)& (1.178, 5.098)&(1.412, 5.149)& (1.600, 5.196)\\
 \hline
 $0.998$ & (0, 4.830)& (1.122, 4.942)& (1.511, 5.037)&(1.782, 5.121)& (1.996, 5.196)\\
\hline\hline
\end{tabular}
\end{center}
\caption{The topmost coordinates ($x_{t}, y_{t}$) of Kerr black hole shadows with the spin parameter $a/M=0.2, 0.4, 0.6, 0.8, 0.998$ and the inclination angle $i=0^{\circ}, 30^{\circ}, 45^{\circ}, 60^{\circ}, 90^{\circ}$ in the celestial coordinate system ($x,y$).}\label{tabsx}
\end{table}

\begin{table}[ht]
\begin{center}
\begin{tabular}{|c|c|c|c|c|c|c|c|c|c|c|}
\hline \hline  \diagbox {$a/M$}{$i(deg)$} & $0^{\circ}$&$30^{\circ}$ &$45^{\circ}$&$60^{\circ}$&$90^{\circ}$ \\
\hline
 $0.2$ & (0,0)& (0.200,0)& (0.284,0)&(0.348,0)& (0.401,0)\\
 \hline
 $0.4$ & (0,0)& (0.407,0)& (0.575,0)&(0.703,0)& (0.810,0)\\
 \hline
 $0.6$ & (0,0)& (0.626,0)& (0.882,0)&(1.077,0)& (1.239,0)\\
 \hline
 $0.8$ & (0,0)& (0.875,0)& (1.229,0)&(1.495,0)& (1.713,0)\\
 \hline
 $0.998$ & (0,0)& (1.226,0)& (1.766,0)&(2.157,0)& (2.443,0)\\
\hline\hline
\end{tabular}
\end{center}
\caption{The center coordinates ($x_{c}, y_{c}$) of Kerr black hole shadows with the spin parameter $a/M=0.2, 0.4, 0.6, 0.8, 0.998$ and the inclination angle $i=0^{\circ}, 30^{\circ}, 45^{\circ}, 60^{\circ}, 90^{\circ}$ in the celestial coordinate system ($x,y$).}\label{tabc}
\end{table}

\section{Determination of the Kerr spin parameter and the inclination angle by the relativistic images in black hole image}

\begin{figure}[ht]
\center{\includegraphics[width=15cm ]{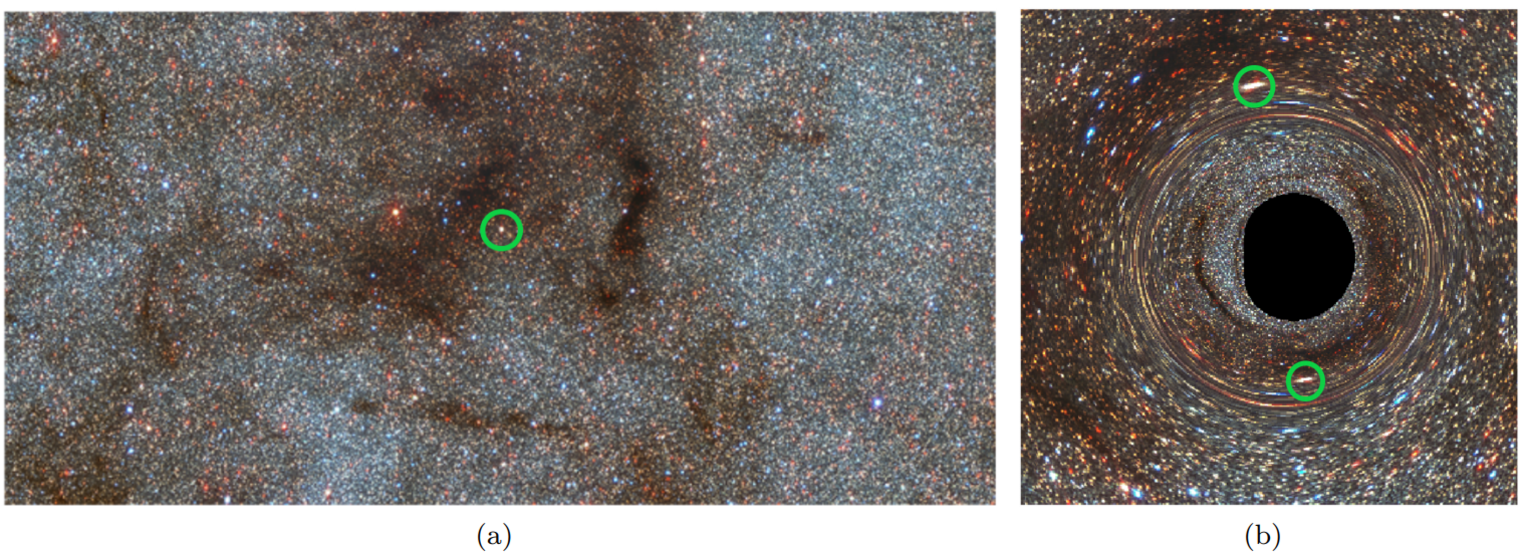}
\caption{(a)The background light source we set, a photo looking toward the center of the Milky Way galaxy taken by the Cerro-Tololo Inter-American Observatory\cite{glx}. (b)Kerr black hole image with the spin parameter $a=0.998M$ and the observer's inclination angle $i=90^{\circ}$.}
\label{bj}}
\end{figure}

\begin{figure}[ht]
\center{\includegraphics[width=12cm ]{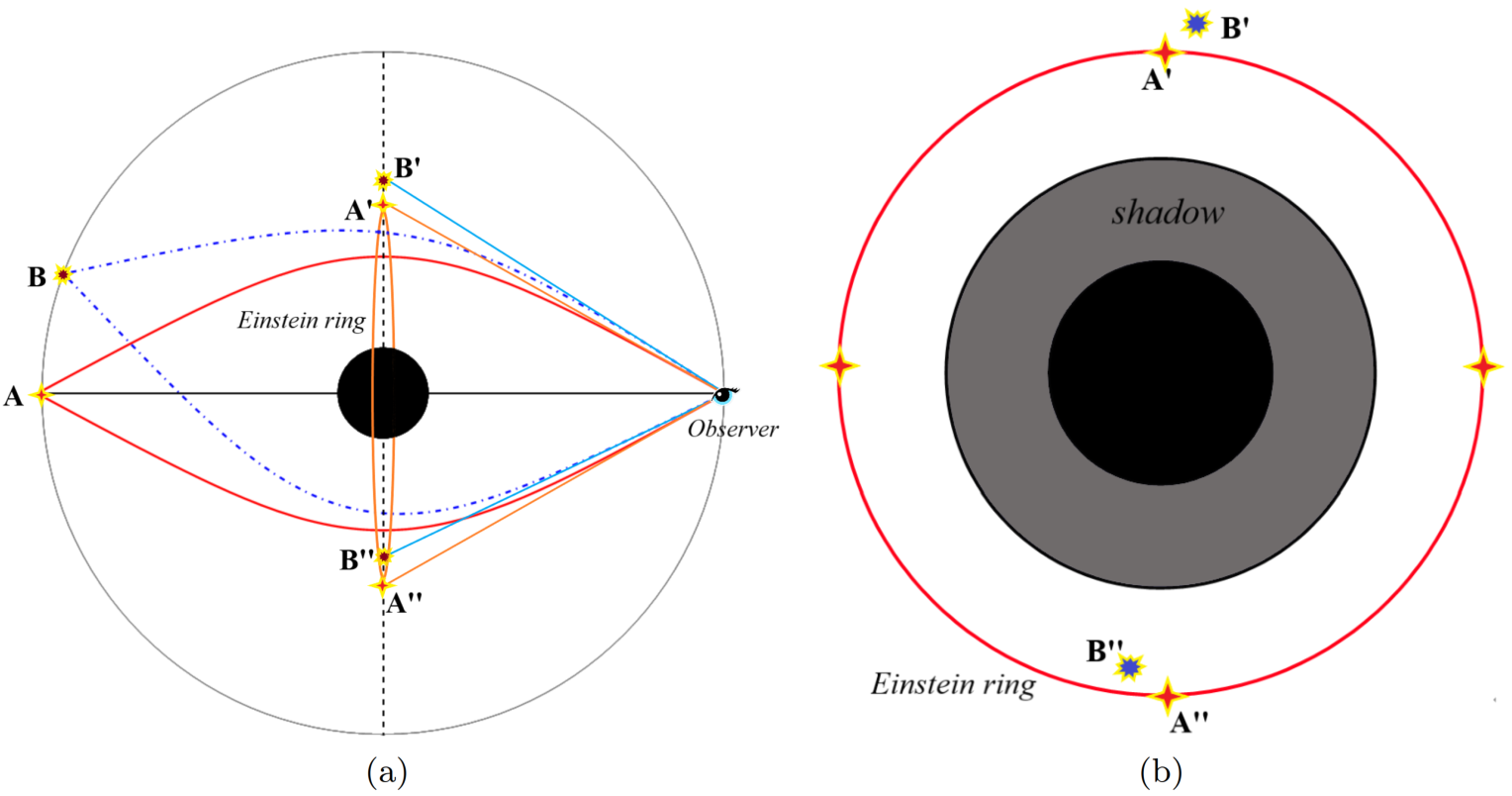}
\caption{The exhibitions of gravitational lensing and the PIs and SIs in black hole image.}
\label{gl}}
\end{figure}

Affected by strong gravitational lensing, an observer may observe multiple images of a luminous celestial object called relativistic images. Actually, there are two infinite sets of relativistic images on either side of the black hole shadow, but only two of them are easily detected, namely primary image (PI) and secondary image (SI). The photons forming PI and SI travel less than a loop around black hole, so they are the outermost relativistic images. The photons forming the other relativistic images could travel complete loops that are near the photon sphere, so these images are concentrated near the shadow boundary and can not be distinguished. We set Fig.\ref{bj}(a) as the background light source which is a photo looking toward the center of the Milky Way galaxy\cite{glx}, and calculated Kerr black hole image by the backward ray-tracing method\cite{sw,swo,astro,chaotic,my,sMN,lf,swo7,mbw,mgw,scc} with the spin parameter $a=0.998M$, the inclination angle $i=90^{\circ}$ in Fig.\ref{bj}(b). In backward ray-tracing method, the light rays are assumed to evolve from the observer backward in time. The light rays falling down into even horizon of black hole correspond to black hole shadow; the light rays arriving at the background light sources correspond to the images of the luminous objects. There is a luminous object inside the green circle in Fig.\ref{bj}(a), and its PI and SI caused by the gravitational lensing are marked inside the green circles in Fig.\ref{bj}(b). The other relativistic images of it can not be distinguished in Fig.\ref{bj}(b). Fig.\ref{gl} exhibits the gravitational lensing, in which the dark disk denotes black hole, $A$ and $B$ denote two luminous objects in the background light source. The lights from $A$ or $B$ can reach the observer along two different paths shown in Fig.\ref{gl}(a), so the observer will see the primary image $B'$ whose lights travel less than half a loop around the black hole, and the secondary image $B''$ whose lights travel more than half a loop. PI and SI are separated by the shadow on either side. In this case, the lights from the luminous object $A$ have an infinite number of paths distributing around the black hole to observer, which results in a ring image of $A$ (the orange ring in Fig.\ref{gl}(a)), namely Einstein ring. Fig.\ref{gl}(b) shows the black hole image, in which the dark and gray disk are black hole and its shadow, respectively. The PI $B'$ is located further away from the black hole shadow than the SI $B''$, and one is situated outside of the Einstein ring(red circle), the other is inside. But it must be mentioned that the light source $A$ of this Einstein ring has the same radius $r_{s}$ with the luminous object $B$. So in black hole images, the PI is the one that is farther from the black hole shadow, and the secondary image is the one that is closer to the shadow.

Black hole shadow carrys some essential signatures about the black hole space-time, so do the relativistic images of luminous celestial objects around the black hole shadow. After calculation, we found that the position between PI and SI depend on the spin parameter $a$ of Kerr black hole and the inclination angle $i$. But locating the positions of the relativistic images need a new celestial coordinates. Because the origin of the celestial coordinates ($x,y$)(\ref{ccxd}) cannot be identified due to the rightward shift of the Kerr black hole shadow as the spin parameter $a$ increases\cite{sha2}. The shadow center ($x_{c}, y_{c}$) can easily be identified, so we defined a new celestial coordinates ($x', y'$) whose origin is the center of black hole shadow as shown in Fig.\ref{xzb}. The new celestial coordinates ($x', y'$) can be expressed by the coordinate ($x, y$) as
\begin{eqnarray}
\label{bhxy}
x'=x-x_{c},\;\;\;\;\;\;\;\;\;\;\;y'=y-y_{c},
\end{eqnarray}
and shares the same scale as the coordinate ($x, y$). Actually, in the astronomical observation the scale of the new celestial coordinates ($x', y'$) is completely flexible.

\begin{figure}[ht]
\center{\includegraphics[width=16cm ]{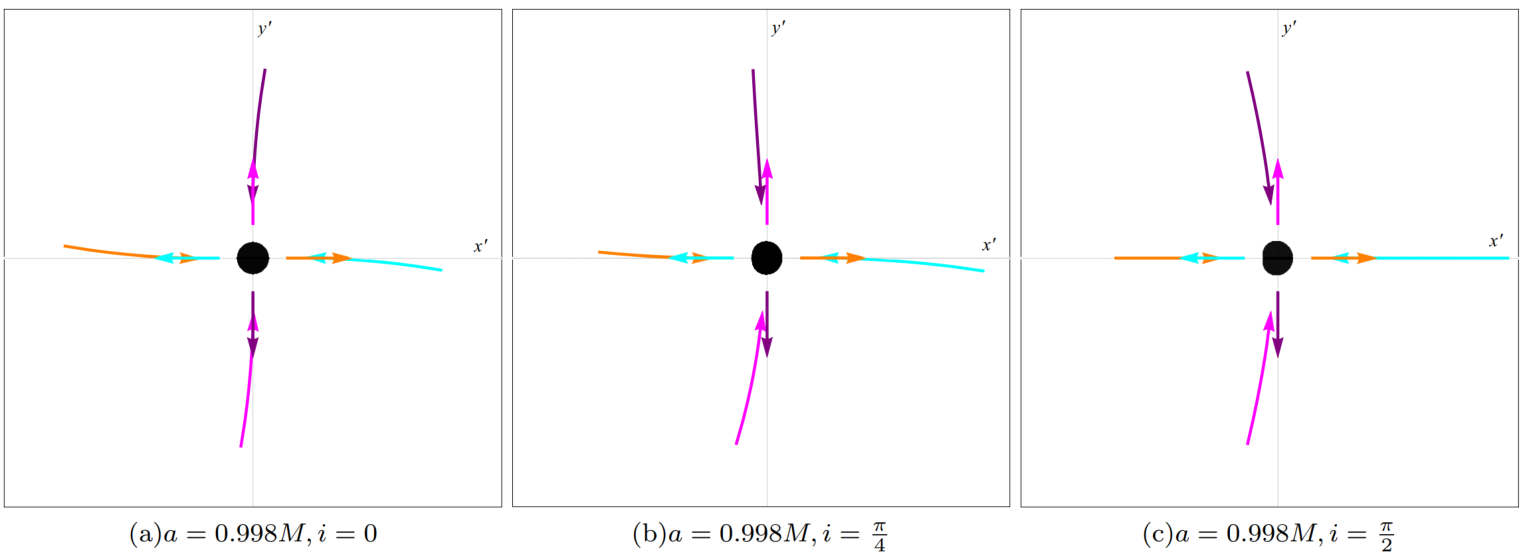}
\caption{Four couples of PI and SI (arrows) in Kerr black hole shadows (black) with the spin parameter $a=0.998M$ and the inclination angle $i=0^{\circ}, 45^{\circ}, 90^{\circ}$ respectively. Here $r_{s}=r_{o}=200M$.}
\label{k04590}}
\end{figure}

Fig.\ref{k04590} presents four couples of PI and SI around Kerr black hole shadows (black) with the spin parameter $a=0.998M$, the inclination angle $i=0^{\circ}, 45^{\circ}, 90^{\circ}$. The four SIs are the arrow starts from ($0, 10$) to ($0, 30$) on the $y'$ axes, the arrow starts from ($0, -10$) to ($0, -30$), the arrow starts from ($-10, 0$) to ($-30, 0$) on the $x'$ axes and the arrow starts from ($10, 0$) to ($30, 0$). By using the backward ray-tracing method, the positions of the light sources can be determined, where we set they locate at $r_{s}=200M$, and the observer locates at $r_{o}=200M$. Thus, the other visible images, PIs, can be obtained, which are marked with the same color as the corresponding SIs, as shown in Fig.\ref{k04590}. The PIs are bigger and have some deformations comparing with the SIs. One can find the closer the image point in SI is to the shadow, the farther away the corresponding image point in PI is from the shadow. And the further away PI is from the black hole shadow, the larger deformation of the PI. This means the closer the SI is to the shadow, the larger deformation of the PI. But if the SI is too close to the shadow it will become more difficult to distinguish due to the image compression near the shadow.

\begin{figure}[ht]
\center{\includegraphics[width=16cm ]{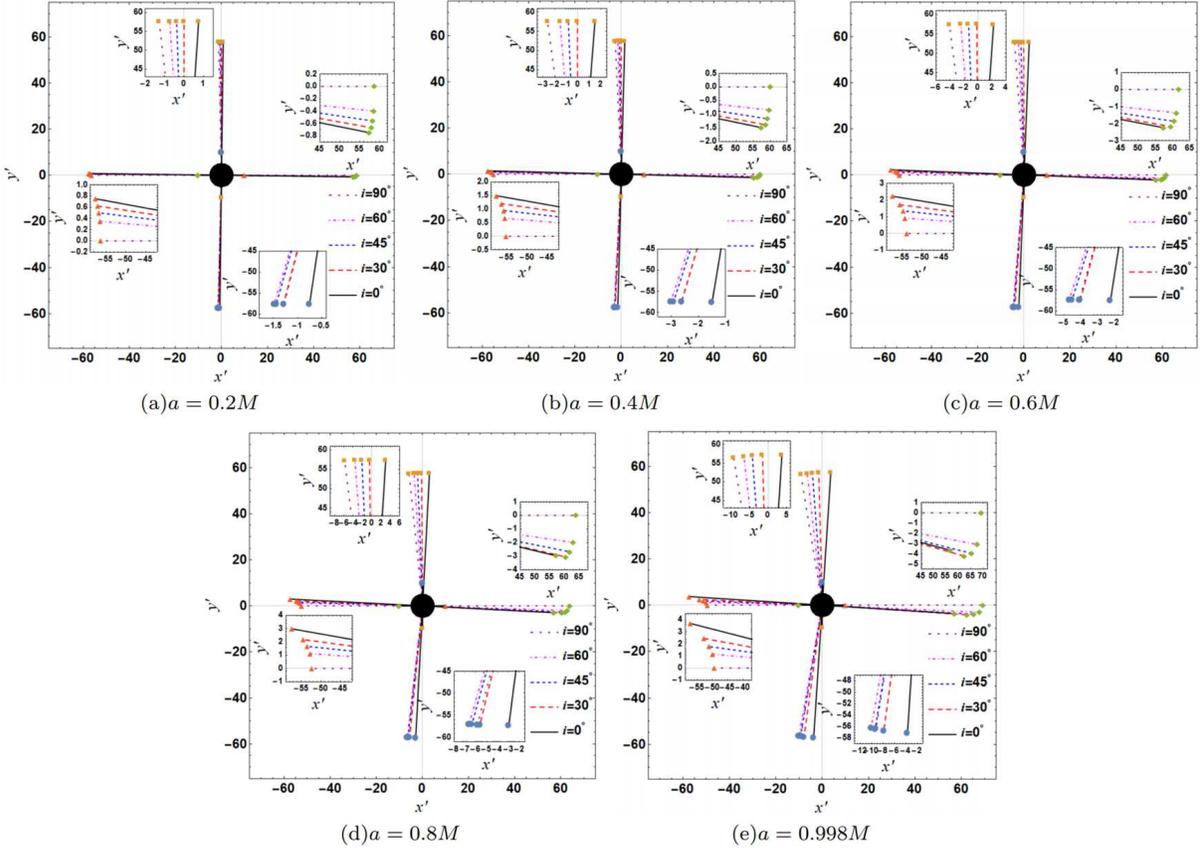}
\caption{The PIs and SIs of four luminous objects (circle, square, rhombus and triangle) in Kerr black hole images with the observer's inclination angle $i=0^{\circ}, 30^{\circ}, 45^{\circ}, 60^{\circ}, 90^{\circ}$, and the spin parameter $a/M=0.2, 0.4, 0.6, 0.8, 0.998$ respectively. Here we set $r_{s}=r_{o}=200M$.}
\label{d298}}
\end{figure}

\begin{figure}[ht]
\center{\includegraphics[width=12cm ]{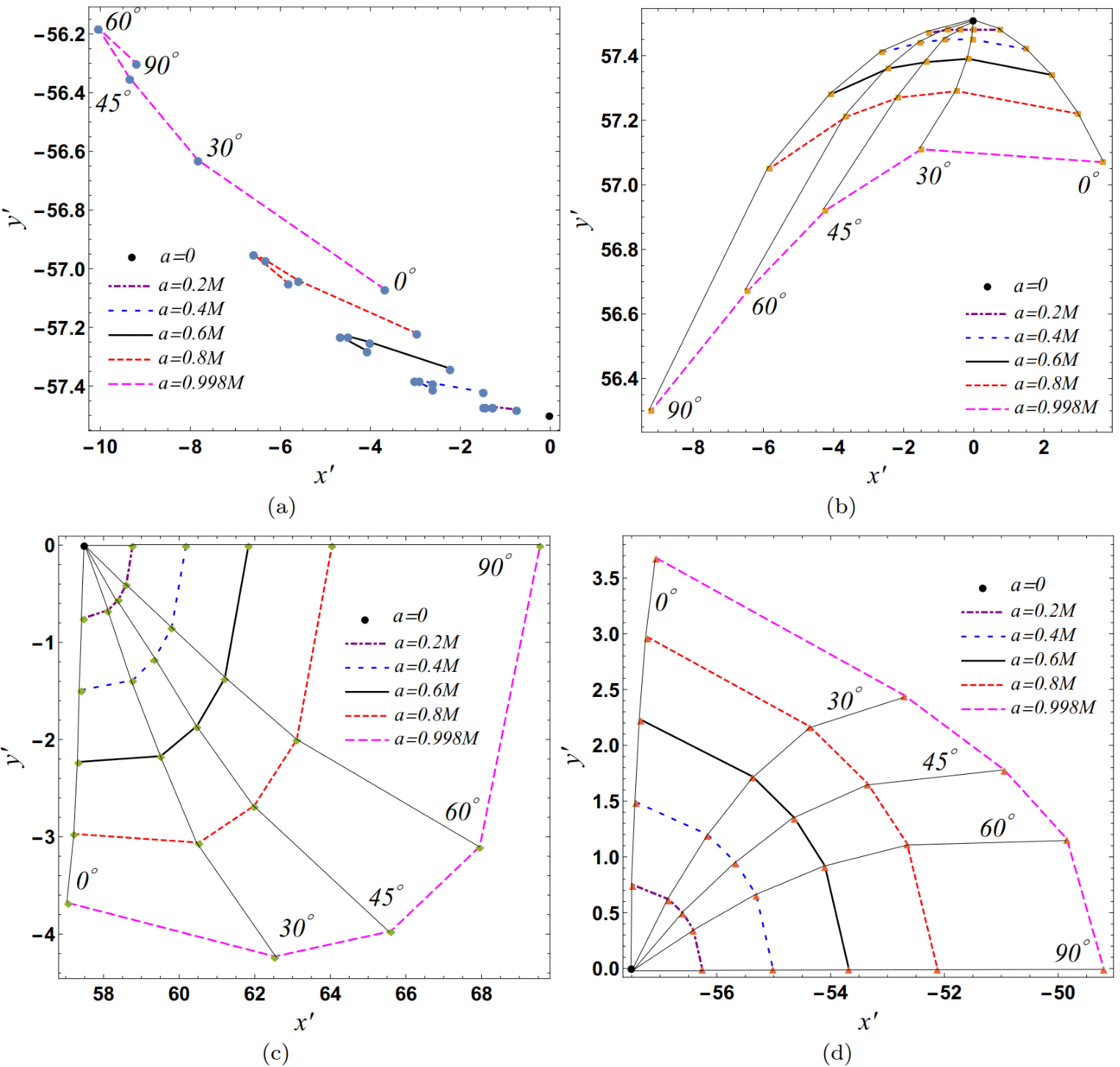}
\caption{The PIs corresponding to the SIs ($0,10$), ($0,-10$), ($-10,0$) and ($10,0$), respectively, under the different spin parameter $a$ and inclination angle $i$.}
\label{dsxzy}}
\end{figure}

In the actual astronomical environment, the light sources are some luminous celestial objects, which do not usually have the same $r_{s}$. According to Fig.\ref{k04590}, we choose four fixed SIs, the round on ($0,10$), the square on ($0,-10$), the rhombus on ($-10,0$) and the triangle on ($10,0$), exhibited in Fig.\ref{d298}. The corresponding PIs also are exhibited in Kerr black hole images with the different spin parameter $a$ and inclination angle $i$ in Fig.\ref{d298}. Here we also set $r_{s}=r_{o}=200M$. The corresponding PIs are on another side of black hole shadow, and are farther away from shadow. From Fig.\ref{d298}, one can find that due to the rotation of the black hole, the position of the PIs will change with the inclination angle $i$, and this alterations will be amplified as the spin parameter $a$ increases. To illustrate the positional deviations more clearly, we show the PIs corresponding to the SIs ($0,10$), ($0,-10$), ($-10,0$) and ($10,0$), respectively, under the different $a$ and $i$ in Fig.\ref{dsxzy}(a-d). One can find the positions of PIs for each SI are different with the different $i$ except for $a=0$. And as $a$ increases, this positional deviations of PIs increase. So the relative position between PI and SI could be used to determine the spin parameter $a$ and the inclination angle $i$.

\begin{figure}[ht]
\center{\includegraphics[width=16.5cm ]{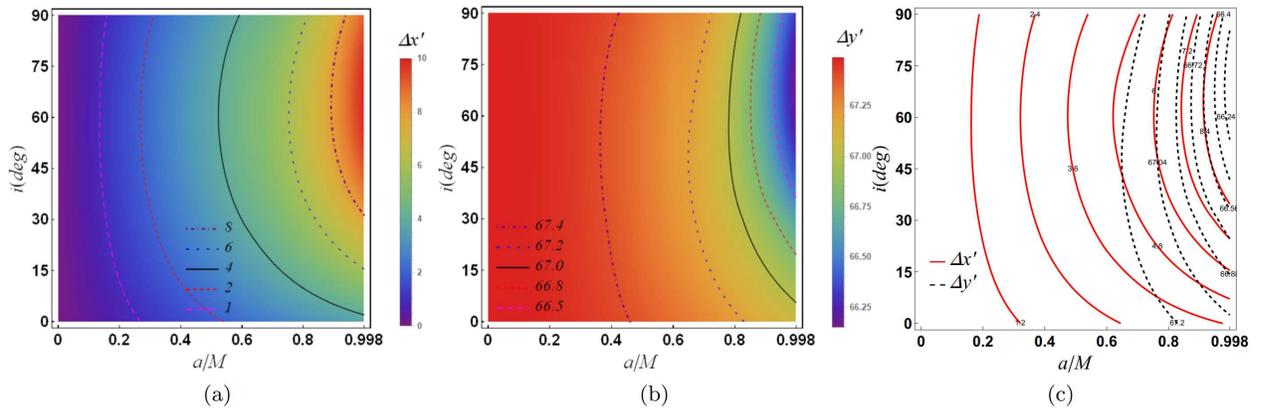}
\caption{The contour maps and lines of $\Delta x'$ and $\Delta y'$ between the SI ($0,10$) and corresponding PIs in terms of $a$ and $i$.}
\label{sd}}
\end{figure}

\begin{figure}[ht]
\center{\includegraphics[width=16.5cm ]{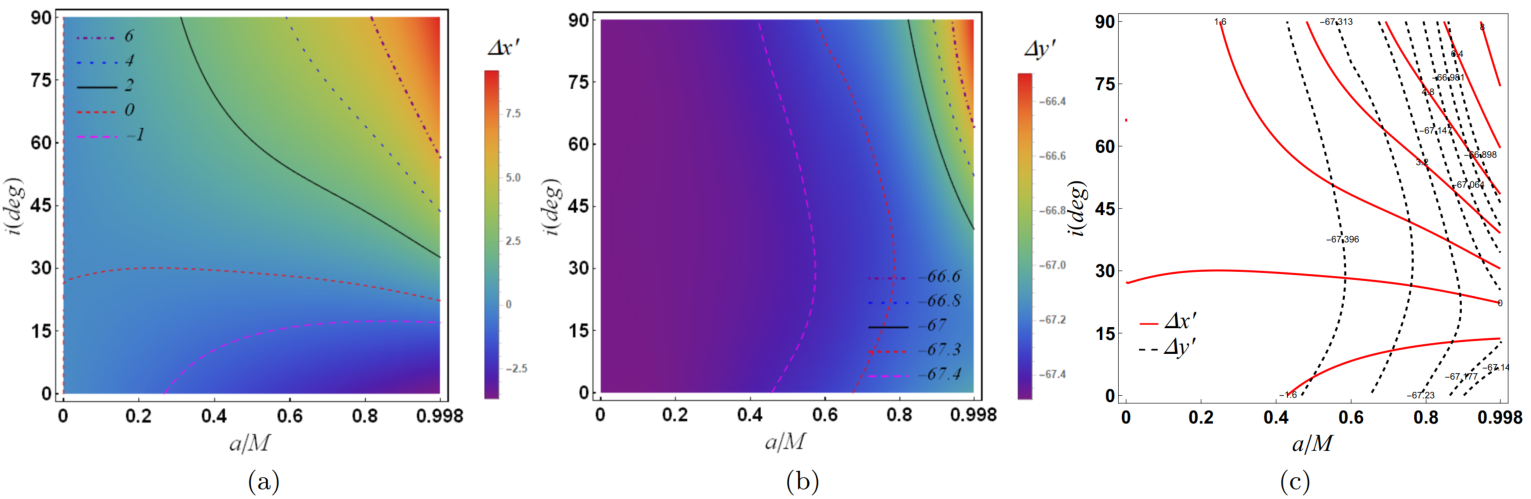}
\caption{The contour maps and lines of $\Delta x'$ and $\Delta y'$ between the SI ($0,-10$) and corresponding PIs in terms of $a$ and $i$.}
\label{xd1}}
\end{figure}

\begin{figure}[ht]
\center{\includegraphics[width=16.5cm ]{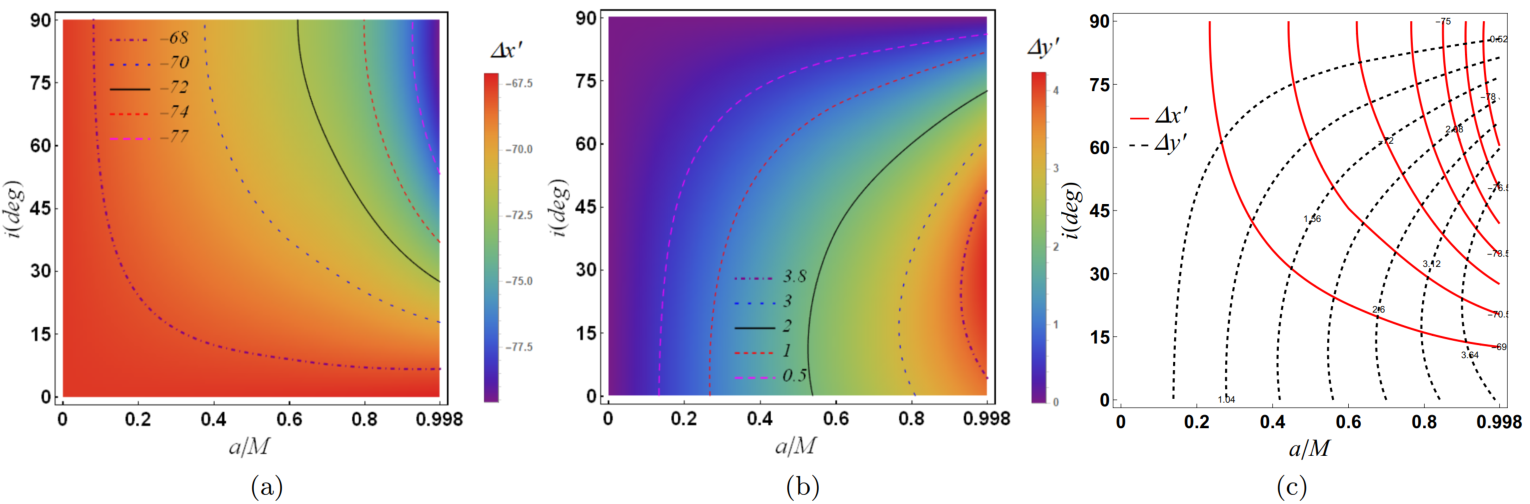}
\caption{The contour maps and lines of $\Delta x'$ and $\Delta y'$ between the SI ($-10,0$) and corresponding PIs in terms of $a$ and $i$.}
\label{zd}}
\end{figure}

\begin{figure}[ht]
\center{\includegraphics[width=16.5cm ]{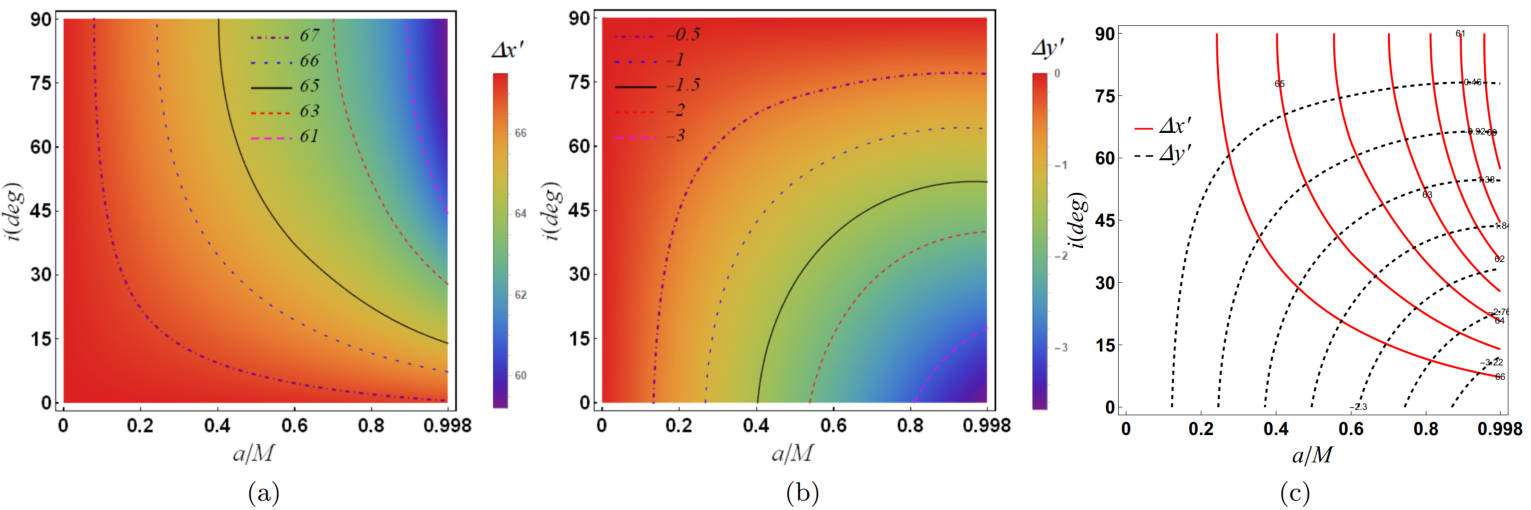}
\caption{The contour maps and lines of $\Delta x'$ and $\Delta y'$ between the SI ($10,0$) and corresponding PIs in terms of $a$ and $i$.}
\label{yd}}
\end{figure}

The relative position between PI and SI can be expresses as
\begin{eqnarray}
\label{delta}
\Delta x'=x'_{s}-x'_{p},\;\;\;\;\;\;\;\;\;\;\;\Delta y'=y'_{s}-y'_{p},
\end{eqnarray}
where ($x'_{p},y'_{p}$) and ($x'_{s},y'_{s}$) are the coordinates of PI and SI in ($x', y'$) coordinate system. We exhibit the contour maps of $\Delta x'$ and $\Delta y'$ between the SI ($0,10$) and corresponding PIs in terms of $a$ and $i$ in Fig.\ref{sd}(a) and (b). One can find $\Delta x'=0$ when $a=0$, and $\Delta x'$ increases, $\Delta y'$ decreases as $a$ increases. With the increase of $i$, $\Delta x'$ first increases and then decreases, and $\Delta y'$ first decreases and then increases. Unfortunately, the contour lines of these $\Delta x'$ and $\Delta y'$ shown in Fig.\ref{sd}(c) reveal that some couples of ($\Delta x', \Delta y'$) can not uniquely determine $a$ and $i$. Fig.\ref{xd1}(a) and (b) show the contour maps of $\Delta x'$ and $\Delta y'$ between the SI ($0,-10$) and corresponding PIs in terms of $a$ and $i$. One can find there is a line with $\Delta x'=0$ (the red dashed line) in Fig.\ref{xd1}(a). Above it $\Delta x'$ increases as $a$ increases, and below it $\Delta x'$ decreases. $\Delta x'$ increases as $i$ increases. In Fig.\ref{xd1}(b), $\Delta y'$ increases as $a$ increases, and first decreases and then increases as $i$ increases. The contour lines of these $\Delta x'$ and $\Delta y'$ are shown in Fig.\ref{xd1}(c), it illustrates the one-to-one correspondence between ($\Delta x'$, $\Delta y'$) and ($a$, $i$). So $\Delta x'$ and $\Delta y'$ between the SI ($0,-10$) and corresponding PI can determine the Kerr spin parameter $a$ and inclination angle $i$ by this contour maps. Furthermore, The span of $\Delta x'$ is $|9.2-(-3.68)|=12.88>2R_{s}\approx10.4$. it is larger than the diameter of Kerr black hole shadow, can be distinguished more easily. But the span of $\Delta y'$ only is $|-66.3-(-67.49)|=1.19$. Fig.\ref{zd} shows the contour maps of $\Delta x'$ and $\Delta y'$ between the SI ($-10,0$) between corresponding PIs in terms of $a$ and $i$. It shows $\Delta x'$ decreases as $a$ and $i$ increase, $\Delta y'$ increases as $a$ increases and decreases as $i$ increases. Fig.\ref{zd}(c) shows the one-to-one correspondence between ($\Delta x'$, $\Delta y'$) and ($a$, $i$), so it also can determine $a$ and $i$. The spin of $\Delta x'$ is $|-67.07-(-79.63)|=12.56$, and the span of $\Delta y'$ is $|4.23-0|=4.23$. Fig.\ref{yd} shows the contour maps of $\Delta x'$ and $\Delta y'$ between the SI ($10,0$) and corresponding PIs in terms of $a$ and $i$. It shows $\Delta x'$ decreases as $a$ and $i$ increase, $\Delta y'$ decreases as $a$ increases and increases as $i$ increases. $\Delta x'$ and $\Delta y'$ between the SI ($10,0$) and corresponding PI also can determine $a$ and $i$ as shown in Fig.\ref{yd}(c). The spin of $\Delta x'$ is $|67.49-59.19|=8.3$, and the span of $\Delta y'$ is $|0-(-3.68)|=3.68$. The span of $\Delta x'$ and $\Delta y'$ between PI and SI are larger enough to give a more accurate $a$ and $i$. In astronomical observation, one even could use the PIs and SIs of multiple light sources to jointly determine Kerr spin parameter $a$ and the inclination angle $i$.

For a given Kerr black hole image with relativistic images, the specific approach to measure the spin parameter $a$ and the inclination angle $i$ by PIs and SIs is:
\begin{enumerate}[(1)]
\item First, the distances of the black hole and the light sources from the observer must be measured, i.e., $r_{o}$ and $r_{s}$. Then locating the center of black hole shadow as the origin of the new celestial coordinate system ($x', y'$). Determining the coordinates ($x'_{p}, y'_{p}$) of PIs and ($x'_{s}, y'_{s}$) of SIs in black hole image, where the scale of this coordinate is completely flexible.
\item Calculating the coordinates ($x_{p}|_{(a,i)}, y_{p}|_{(a,i)}$) of PIs and ($x_{s}|_{(a,i)}, y_{s}|_{(a,i)}$) of SIs in the celestial coordinates system ($x, y$) (\ref{ccxd}) with different spin parameter $a$ and inclination angle $i$, which is given by
\begin{eqnarray}
\label{bh}
x|_{(a,i)}=\frac{(x_{r}-x_{l})|_{(a,i)}}{x'_{r}-x'_{l}}x'+x_{c}|_{(a,i)},\;\;\;\;\;\;\;\;\;\;\;y|_{(a,i)}=\frac{y_{t}|_{(a,i)}}{y'_{t}}y',
\end{eqnarray}
where $x_{r}, x_{l}, x_{c}, y_{t}$ are the functions of ($a, i$) as seen in (\ref{xd1w}) and TABLE \ref{tabl}-\ref{tabc}, and $x'_{r}, x'_{l}, y'_{t}$ can be measured in black hole shadow.
\item According to the coordinates ($x_{p}|_{(a,i)}, y_{p}|_{(a,i)}$) of PIs and ($x_{s}|_{(a,i)}, y_{s}|_{(a,i)}$) of SIs, calculating where their light sources are located by the backward ray-tracing method. Find out the value of $a$ and $i$, when PI and SI share a common light source, is the spin parameter $a$ and inclination angle $i$ of the black hole.
\end{enumerate}

\section{Time delays between the primary and secondary images}
\begin{figure}[ht]
\center{\includegraphics[width=12cm ]{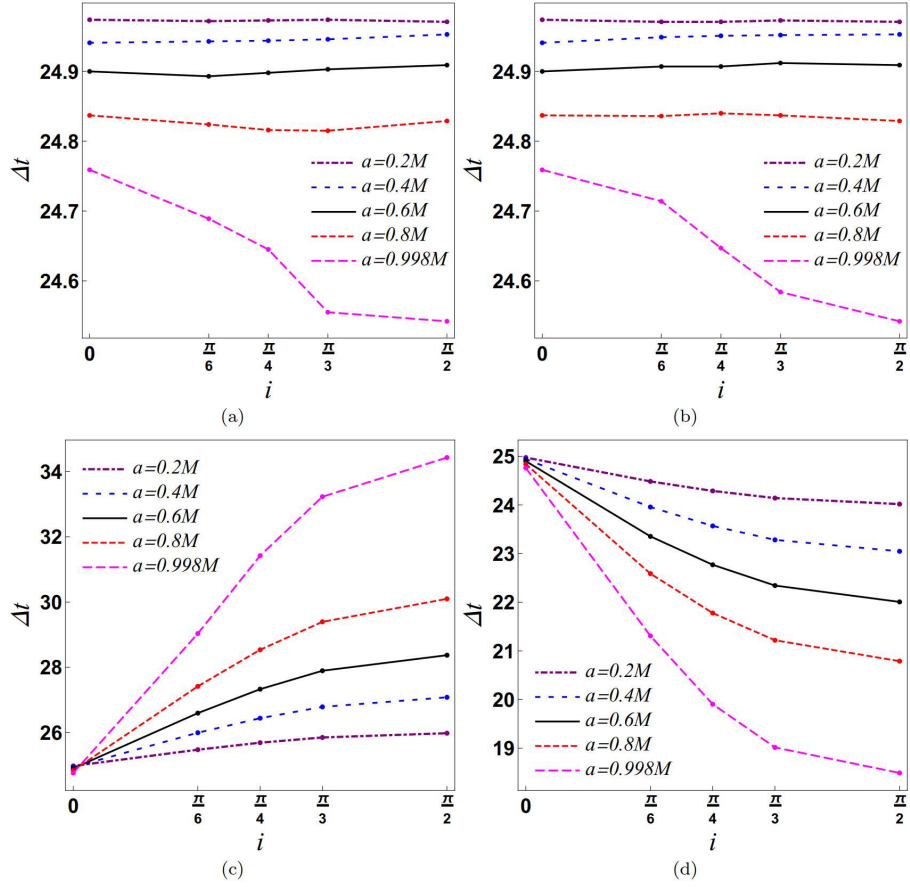}
\caption{(a) The time delays $\Delta t=(t_{s}-t_{p})$ between SI ($0,10$) and corresponding PI under the different spin parameter $a$ and inclination angle $i$. (b) $\Delta t$ between  SI ($0,-10$) and corresponding PI. (c) $\Delta t$ between SI ($-10,0$) and corresponding PI. (d) $\Delta t$ between SI ($10,0$) and corresponding PI. These PIs and SIs are the same as that in Fig.\ref{d298} and Fig.\ref{dsxzy}.}
\label{tsxzy}}
\end{figure}

From Fig.\ref{gl}, One can find the photons forming SI have longer paths than that forming PI. Moreover, the photons forming SI are subject to a greater gravitational field due to the closer distance to black hole. It makes the photons forming the SI cost more time to the observer. So there is a time delay between the PI and SI that the lights reach observer. As mentioned above, the relative position between PI and SI is different under the different spin parameter $a$ and inclination angle $i$. The different paths of these PIs and SIs might cause the different time delays. Fig.\ref{tsxzy} shows the time delays $\Delta t=(t_{s}-t_{p})$ between PIs and SIs under the different $a$ and $i$, where $t_{p}$ and $t_{s}$ are the time for the photon in PI and SI to reach the observer from light source, and PI and SI are the same as that in Fig.\ref{d298} and Fig.\ref{dsxzy}. Fig.\ref{tsxzy} (a) and (b) show the time delays $\Delta t$ between SI ($0,10$) and corresponding PI, and between SI ($0,-10$) and corresponding PI, respectively. Since these PIs and SIs are almost on the $y'$-axis, the time delay $\Delta t$ between them barely change with $i$ for fixed $a$, but for $a=0.998M$, it decreases as $i$ increases. For fixed $i$, the time delays $\Delta t$ decreases as $a$ increases. Fig.\ref{tsxzy} (c) and (d) show the time delays $\Delta t$ between SI ($-10,0$) and corresponding PI, and between SI ($10,0$) and corresponding PI, respectively. Affected by the dragging effect, the time delays $\Delta t$ between SI ($-10,0$) and corresponding PI increases as $a$ increases. Since PI and SI are almost on the $x'$-axis, the increase in $i$ means that they are getting closer to the black hole equator. Which leads to the gradual enhancement of the dragging effect, and thus the increase of $\Delta t$. On the contrary, the time delays $\Delta t$ between SI ($10,0$) and corresponding PI decreases as $a$ and $i$ increases. The time delays $\Delta t$ between PIs and SIs in conjunction with the relative positions of them could allow us to measure the spin parameter $a$ and inclination angle $i$ of Kerr black hole more precisely.

\section{Conclusion}

We studied the relativistic images caused by strong gravitational lensing in Kerr black hole images, which carry some essential signatures about the black hole space-time. In order to locate the relativistic images, we defined a new celestial coordinates whose origin is the center of black hole shadow. Under the influences of the dragging effect caused by rotating black hole and the inclination angle of observer, the relative positions between PIs and SIs are different under the different Kerr spin parameter $a$ and inclination angle $i$. We marked four SIs above, below, on the left and on the right of the shadow center respectively, and found the positions of the PIs are different for different $a$ and $i$. In astronomical observation, the relative position between PI and SI in black hole image can be used to determine the Kerr spin parameter $a$ and the inclination angle $i$. Furthermore, we provided the specific approach to measure the value of $a$ and $i$ by the relativistic images. The time delays between the PIs and SIs are different under the different $a$ and $i$. The time delays in conjunction with the relative positions between PIs and SIs could allow us to measure the value of $a$ and $i$ more precisely. These relativistic images are as unique as fingerprints for black hole space-time, by which one can further determine other parameters of all kinds of compact objects, such as Kerr-like black holes and so on, and verify various theories of gravity. Our results provide a new method to implement parameter estimation in the study of black hole physics and astrophysics.

The black hole pictures observed by EHT Collaboration indicate it is the image of accretion disk around black hole shadow. The accretion disk is the light source close to black hole. Further research is needed on whether the parameters of black holes and the inclination angle can be extracted from the accretion disk image. In addition, what is the image of naked singularity looks like? Whether the parameters of naked singularity can be extracted from the images? To research the naked singularity images could help us to further test the cosmic censorship hypothesis. The study of the relativistic images in astronomical images could provide a very important help for gravitation theory.

\section{\bf Acknowledgments}

This work was supported by the National Natural Science Foundation of China under Grant No. 12105151, the Shandong Provincial Natural Science Foundation of China under Grant No. ZR2020QA080, and was partially supported by the National Natural Science Foundation of China under Grant No. 11875026, 11875025 and 12035005.

\end{document}